\documentclass[11pt,a4paper]{article} 
\pdfoutput=1
\usepackage{jcappub}
\usepackage{enumerate}
\usepackage{epsfig} 
\usepackage{float} 
\usepackage[caption = false]{subfig}
\usepackage{wasysym} 
\usepackage{mathrsfs} 
\usepackage{amsfonts} 
\usepackage{amsbsy} 
\usepackage{amscd} 
\usepackage{pstricks} 
\usepackage{multirow} 
\usepackage{tikz}
\usepackage{color}
\usepackage{slashed}
\usepackage{array}
\usepackage{tablefootnote} 
\usepackage{enumitem}
\usepackage[utf8]{inputenc}
\usepackage{hyperref}
\usepackage{color}
\definecolor{myred}{rgb}{0.6,0,0} 
\definecolor{myblue}{rgb}{0,0.2,0.4}
\definecolor{mygreen}{rgb}{0,0.9,0.1}
\definecolor{hc}{rgb}{.9,0.1,0.7}
\definecolor{hcout}{rgb}{.9,0.7,0.9}
\definecolor{Orange}{rgb}{0,0.2,0.9}

\makeatletter
\gdef\@fpheader{}

\newcommand{\fmslash}[2][0mu]{%
  \mathchoice
    {\fmsl@sh\displaystyle{#1}{#2}}%
    {\fmsl@sh\textstyle{#1}{#2}}%
    {\fmsl@sh\scriptstyle{#1}{#2}}%
    {\fmsl@sh\scriptscriptstyle{#1}{#2}}}
\newcommand{\fmsl@sh}[3]{%
  \m@th\ooalign{$\hfil#1\mkern#2/\hfil$\crcr$#1#3$}}
\makeatother
\vspace*{0.7cm}
\newcommand{\lsim}{{\;\raise0.3ex\hbox{$<$\kern-0.75em\raise-1.1ex\hbox{$\sim$}}\;}}
\newcommand{\gsim}{{\;\raise0.3ex\hbox{$>$\kern-0.75em\raise-1.1ex\hbox{$\sim$}}\;}}

\usepackage{array}
\newcolumntype{C}[1]{>{\centering\arraybackslash$}p{#1}<{$}}

\usepackage{bm} 
\newcommand{\be}{\begin{equation}}
\newcommand{\ee}{\end{equation}}
\newcommand{\bes}{\begin{equation*}}
\newcommand{\ees}{\end{equation*}}
\newcommand{\bea}{\begin{eqnarray}}
\newcommand{\eea}{\end{eqnarray}}
\newcommand{\beas}{\begin{eqnarray*}}
\newcommand{\eeas}{\end{eqnarray*}}
\usepackage{booktabs} 
\DeclareRobustCommand{\orcidicon}{
	\begin{tikzpicture}
		\draw[lime, fill=lime] (0,0)
		circle [radius=0.16]
		node[white] {{\fontfamily{qag}\selectfont \tiny ID}};
		\draw[white, fill=white] (-0.0625,0.095)
		circle [radius=0.007];
	\end{tikzpicture}
	\hspace{-2mm}
}
\foreach \x in {A, ..., Z}{\expandafter\xdef\csname orcid\x\endcsname{\noexpand\href{https://orcid.org/\csname orcidauthor\x\endcsname}
		{\noexpand\orcidicon}}
}

\title{Constraining eV-scale axion-like particle dark matter: 
insights from the M87 Galaxy}
\author[a\orcidB{}]{\bf Arpan Kar,}
\author[b\orcidC{}]{Sourov Roy,}
\author[b\orcidA{}]{Pratick Sarkar}

\affiliation[a]{Laboratoire de Physique Théorique et Hautes Énergies (LPTHE),CNRS \& Sorbonne Université, 4 Place Jussieu, Paris, France}
\affiliation[b]{School of Physical Sciences, Indian Association for the Cultivation of Science,\\2A \& 2B Raja S.C Mullick Road, Kolkata-700032, India}

\emailAdd{arpankarphys@gmail.com}
\emailAdd{tpsr@iacs.res.in}
\emailAdd{spsps2523@iacs.res.in}

\abstract{Axion-like particles (ALPs) can account for the observed dark matter (DM) of the Universe and if their masses are at the eV scale, they can decay into 
infrared, optical and ultraviolet photons 
{with a decay lifetime larger than the age of the Universe}. 
We analyze multi-wavelength data obtained from the central region of Messier 87 (M87) galaxy 
by several telescopes, such as, Swift, Astrosat, Kanata, Spitzer and 
International Ultraviolet Explorer 
in the infrared to ultraviolet frequencies 
($\sim 2\times10^{14} \, {\rm Hz} - 3\times10^{15}$ Hz), 
to constrain the narrow emission lines indicative of the eV scale ALP DM decay. 
We derive constraints on the ALP coupling to two photons ($g_{a\gamma\gamma}$) for ALP mass range $2 \, {\rm eV} \lesssim m_a \lesssim 20 \, {\rm eV}$, 
assuming ALPs form the DM in the M87 halo. 
We find that our bounds on ALP-two-photon coupling 
can become stronger than the existing ones by an order of magnitude in the ALP 
mass range $8 \, {\rm eV} \lesssim m_a \lesssim 20 \, {\rm eV}$.}
\begin{document}
     \maketitle
	\flushbottom
\section{Introduction}
Numerous astrophysical and cosmological observations firmly support the existence of 
DM, a non-baryonic, minimally interacting, cold matter 
component that provides about 25\% of the energy density of our universe \cite{Planck:2018vyg}. 
However, the microscopic features of DM are still unknown. Extensive efforts in its search suggest that DM resides in the form of new, unknown particles. 

Quantum Chromodynamics (QCD) presents a fundamental challenge known as the 
Strong Charge-Parity (CP) problem, with axions emerging as a potential solution to this issue\cite{Peccei:1977hh,Peccei:1977ur, Weinberg:1977ma, Wilczek:1977pj, Shifman:1979if, Zhitnitsky:1980tq,Kim:1979if, Dine:1981rt}. The non-observation of the neutron electric dipole moment\cite{Pendlebury:2015lrz} constrains the CP- violating $\theta$- parameter to be surprisingly small, $|\theta|\leq 10^{-10}$, while its value could generically be $\mathcal{O}(1)$. A most promising solution to this strong CP problem is to introduce a new global U(1) symmetry which spontaneously breaks at some high-energy scale through the Peccei-Quinn mechanism to produce a pseudo-Nambu-Goldstone boson, known as the axion. Despite the small mass, the axions can be produced non-thermally in the early universe and are stable in cosmological time scales and can contribute substantially to the current fraction of the energy density of the Universe 
in the form of cold DM 
\cite{Preskill:1982cy,Abbott:1982af,Dine:1982ah,KHLOPOV1999105,Duffy:2009ig,Hui:2016ltb, Chadha-Day:2021szb,Adams:2022pbo,Cirelli:2024ssz}. 
The QCD axion couples to Standard Model photons via electromagnetic anomaly and the coupling, $g_{a\gamma\gamma}$, varies linearly with the axion mass $m_a$. 
On the other hand, axion-like particles (ALPs), which posses properties similar to the QCD axion, are, however, more generic in the sense that they do not exhibit a direct one-to-one relationship between their mass and the photon coupling $g_{a\gamma\gamma}$. 
ALPs which naturally arise in low-energy effective theories from higher-dimensional theories \cite{Svrcek:2006yi, Arvanitaki:2009fg, Maity:2007un,Haque:2024dxm} 
can be a potential candidate for the observed DM of the Universe. 
Thus, beyond the well-motivated QCD axion, it is also intriguing to explore the whole parameter space of ALPs. 

Due to its couplings to photons, ALPs can decay in to two photons, although the decay rate should be slow enough so that they are stable on the cosmological time scale to explain the observed DM. Astronomical telescopes can provide a powerful means of searching for signals from such ALP DM decay. The decay of ALP DM particle inside 
a DM dense region such as a galactic DM halo can produce a distinctive monochromatic emission line that can be distinguished from astrophysical backgrounds. 
Over the years, searches for these emission lines have spanned a broad range of frequencies, from radio waves to X-rays \cite{Roach:2022lgo,Bernal:2022xyi,Bessho:2022yyu,Roy:2023omw,PhysRevD.104.023021,Wang:2023imi,Carenza:2023qxh,Porras-Bedmar:2024uql}. 
In particular, if the mass $m_a$ of the ALP particles is on the order of electronvolts (eVs) then their decay would produce photons in the infrared, optical, and ultraviolet bands. In this context, rigorous spectroscopic searches have been performed in the infrared to optical frequency bands \cite{Grin:2006aw,Regis:2020fhw,Caputo:2020msf,Wadekar:2021qae,Nakayama:2022jza,Todarello:2023hdk,Janish:2023kvi,Yin:2024lla,Regis:2024znx} 
which impose constraints on the $g_{a\gamma\gamma} - m_a$ plane for ALPs, further encouraging the exploration of ALPs at the eV scale. Until recently, no searches had been conducted in the far-ultraviolet band, with the exception of a very recent study \cite{todarello2024bounds}. 

In this study, we constrain the line emission from the decay of eV scale ALP DM 
using the optical-ultraviolet bands in spectroscopic data collected by the Neil Gehrels Swift Observatory (which has been operational for 19 years). 
Its Ultraviolet/Optical Telescope (UVOT) is designed to capture optical and UV emissions from astrophysical sources. Additionally, we utilized data from the Kanata Telescope at the Hiroshima Astrophysical Science Center, which operates in the near-infrared and optical bands \cite{2012SPIE.8446E....M}. 
We also explored the near-UV and visible bands using the UV Imaging Telescope (UVIT) onboard the Astrosat mission, which features dedicated near-UV, visible, and far-UV channels \cite{Tandon_2017,2021JApA...42...20G}. 
This work focuses on observations of the central region of Messier 87 (M87) galaxy, 
a DM-rich environment. In such regions, decay of ALP DM particles into two photons could lead to strong emission lines, non-observation of which by current telescopes targeting at the central part of M87 put constraint on the ALP-photon coupling.
Also, the recent imaging of M87* \cite{EventHorizonTelescope:2019uob,EventHorizonTelescope:2019jan,EventHorizonTelescope:2019ths,EventHorizonTelescope:2019pgp,EventHorizonTelescope:2019ggy,Akiyama_2021,EventHorizonTelescope:2021srq}, the supermassive black hole (SMBH) at the center of M87 galaxy, provides an unprecedented opportunity to probe phenomena beyond the Standard Model\cite{Nomura:2022zyy,Roy:2023rjk,Chen:2019fsq}. 
Two major Multi Wavelength EHT campaigns (MWL2017\cite{EventHorizonTelescope:2021dvx} and  MWL2018\cite{EventHorizonTelescope-Multi-wavelengthscienceworkinggroup:2024xhy}), conducted by the 
EHT Multi-wavelength Science Working Group (MWL WG), analyzed multi-band data from millimeter waves to TeV $\gamma$-rays for the central region of M87 galaxy. 
This study leverages data from the 2018 campaign (MWL2018) 
(with a focus on the ultraviolet and optical bands) as well as 
the data from previous observations \cite{Perola1980,Biretta,Despringe,Birettasparks,Marshall:2001de,Perlman2003,Beuther:2006ad,Shi:2006vs,Tan:2006vk,Perlman:2007zc} by 
Spitzer and International Ultraviolet Explorer
focused towards M87 center in the infrared-ultraviolet band 
to constrain the decay of eV scale ALP DM.

This paper is organized as follows: section~\ref{Axion Signal} describes the computation of the ALP DM decay signal; section~\ref{data} introduces the observational dataset; section~\ref{Methodology and  Results} details the methodology used in our analysis and the obtained results; finally in section~\ref{conclusion} we conclude. 

\section{ALP DM decay signal}\label{Axion Signal}

In this work, we look for infrared-optical-ultraviolet emission lines from 
ALP DM decay~\footnote{Our study, however is sensitive to any DM candidate which produces narrow emission signals.}. 
The ALP can decay to two photons via the interaction\cite{Janish:2023kvi,Carenza:2023qxh,Kar:2022ngx}
\begin{equation}
 \mathcal{L}=-\frac{1}{4}g_{a\gamma \gamma}a F_{\mu \nu}\tilde{F}^{\mu \nu}   
\end{equation}
where $a$ represents the ALP field, $F_{\mu \nu}$ and $\tilde{F}^{\mu \nu}$ denote the electromagnetic field strength tensor and its dual, respectively, and 
$g_{a\gamma \gamma}$ is the ALP-photon coupling. 
The ALP decay rate is given by:
\begin{equation}
\Gamma_a = \frac{g_{a\gamma\gamma}^2 m_{a} ^3}{64\pi}
\end{equation} 
where $m_{a}$ is the ALP mass. Note that the lifetime ($1/\Gamma_a$) 
should be larger than the age of the universe so that the ALP can explain 
the observed DM density of the Universe. 

The photon flux density, i.e., the radiated power per unit area per unit frequency, from the decay of ALP DM is given by \cite{Caputo:2018vmy}: 
\begin{equation}
S_{decay}=\frac{\Gamma_{a} }{4 \pi} \, \int d\theta \, 2\pi \sin\theta \, 
\int_{l.o.s} dl \,\, \frac{\rho_{a}(r(l, \theta))}{\Delta \nu(r(l,\theta))} \, .
\label{ALP_signal}
\end{equation}
Here $\Delta \nu$ is the width of the produced photon line, given by 
$\Delta \nu = \nu_c \, \sigma_d/c$, where 
$\nu_{c}$ is the central frequency (= $m_a / 4\pi$) of the 
photon line, $\sigma_d$ is the velocity dispersion of DM particles 
within the DM halo and $c$ is the speed of light. 
The decay signal will follow a Gaussian distribution, centered around the flux $S_{decay}$ as the peak value, with $\Delta \nu$ representing the spread of the signal. 
We study ALP DM decay signal from the central region of the M87 galaxy which 
hosts a super massive black hole (SMBH). 
Around the SMBH the velocity dispersion of DM is given by 
$\sqrt{\frac{2G_N M_{BH}}{r}} $ \cite{Edwards:2019tzf}, 
where $G_N$ is the gravitational constant 
and $M_{BH}$ is the black hole mass which for the 
central SMBH in M87 is $M_{BH} \simeq 6.5\times 10^{9} M_{\odot}$ \cite{EventHorizonTelescope-Multi-wavelengthscienceworkinggroup:2024xhy}. 
Additionally, there is also an outer halo component of the DM velocity dispersion 
which is of the order of $10^{-3}$c. 
The integral in Eq. \ref{ALP_signal} is performed over the line of sight (l.o.s) $l$ 
between the emitting object and the location of Earth and 
the angle $\theta$ covered by the observing telescope. 
The radial distance from the center of M87 galaxy $r (l, \theta)$ is given 
by $r = \sqrt{l^2 + d^2 -2ld\cos\theta}$, where $d$ is the distance of the 
M87 center from Earth. 
In order to take into account the effect of absorption of the DM induced photons,  
we remove the contribution from the central region 
$r \lesssim 100 R_{s}$ (with $R_{s}$ being the Schwarzchild radius of the SMBH) from Eq.~\ref{ALP_signal},
where the absorption can be stronger \cite{EventHorizonTelescope-Multi-wavelengthscienceworkinggroup:2024xhy}.

In Eq.~\ref{ALP_signal}, $\rho_{a} (r)$ describes the DM mass density distribution in the 
inner region of M87 galaxy. For $\rho_{a}(r)$ we adopt the Navarro-Frenk-White (NFW) profile 
\cite{Navarro:1995iw, HAWC:2023bti},
\begin{equation}
\rho_{\rm NFW}(r)=\frac{\rho_{0}}{ (r/r_{0}) (1+r/r_{0})^{2}} \, ,
\label{eq:NFW_profile}
\end{equation}
with $\rho_{0}$ and $r_{0}$ are the scale density and scale radius, respectively. 

The parameters for the M87 galaxy within the Virgo Cluster are: 
$\rho_{0} = 6.96\times 10^{5} \, M_{\odot}/{\rm kpc^{3}}$, $r_{0} = 403.8$ kpc, 
spread of the virial radius of the DM halo is approximately 1.7 Mpc and 
distance from the earth $d = 17.2$ Mpc \cite{HAWC:2023bti}. 
Additionally, we will consider the NFW DM profile from \cite{Lacroix:2015lxa} 
with different profile parameters as well as a cored halo profile from \cite{DeLaurentis:2022nrv}, which corresponds to the Burkert profile. 

In the bottom panel of Fig.~\ref{fig:The_Data_Plot}, the blue line shows an 
example of the ALP DM signal from the central region of M87. Such a signal is 
estimated with the DM profile parameters mentioned above. 
The ALP mass is considered to be $m_{a}=9.51$ eV, with the value of 
$g_{a\gamma\gamma}$ set to $1.52\times 10^{-11} \, \rm GeV^{-1}$ which 
corresponds to the existing bound on the coupling at this mass. 
This $m_{a}$ produces an observed frequency of $\nu = 1.15 \times 10^{15}$ Hz. 
The angular width of observation is taken to be the one 
that corresponds to the resolution used at that frequency. 
 
In the literature, the DM density near the center of galaxies (such as M87) is often 
described by a spike profile (due to the presence of a SMBH at the center) 
\cite{Gondolo:1999ef,Lacroix:2015lxa,Balaji:2023hmy,Phoroutan-Mehr:2024cwd}. 
However, for this study, we only adopt the canonical NFW profile. 
We checked that our DM decay signal remains almost unaffected even when a spike profile is superimposed on the NFW profile.

\section{Observations and Data}\label{data}

\begin{figure*}[ht!]
\centering
\includegraphics[scale=0.5]{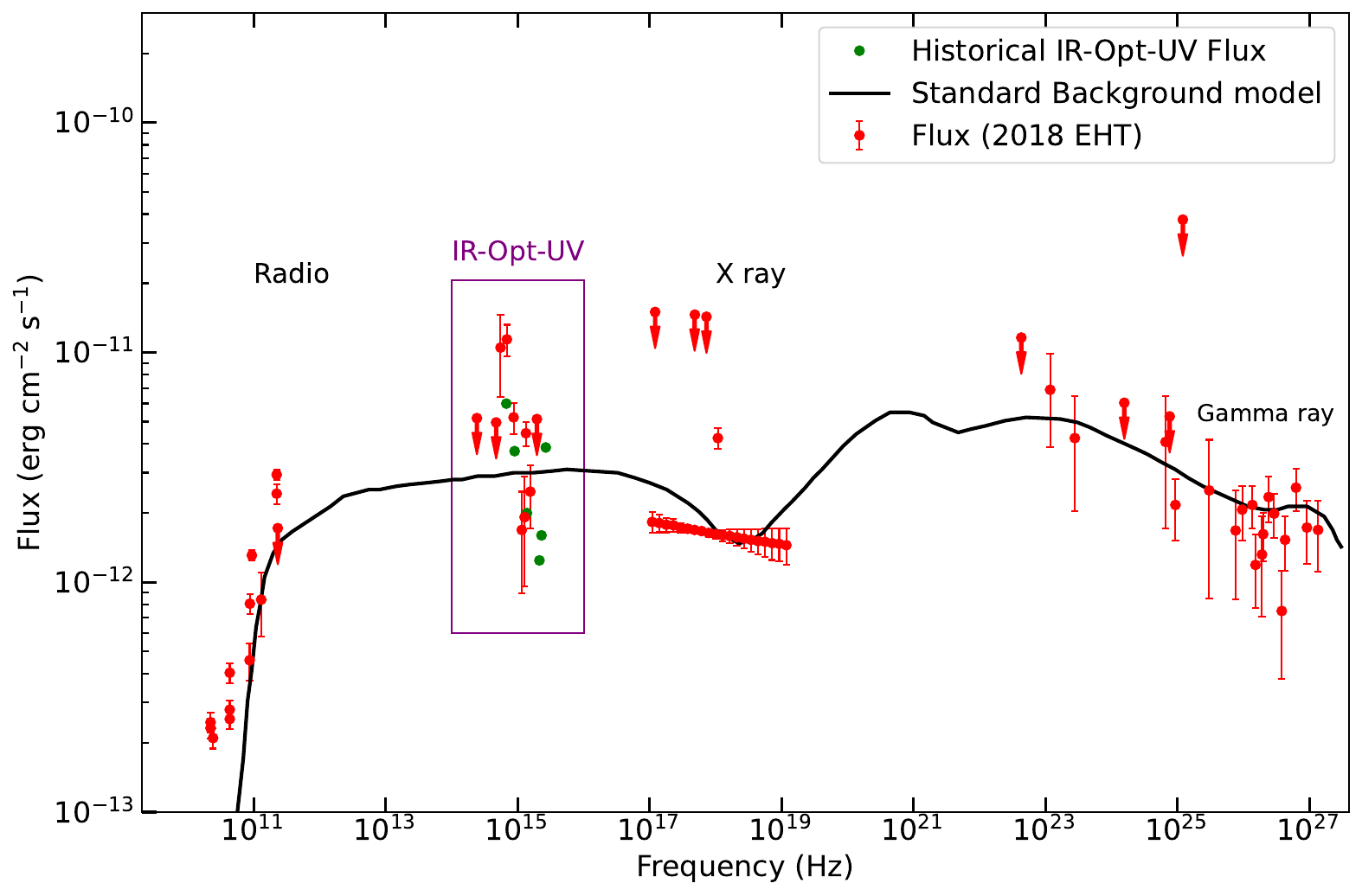}
\includegraphics[scale=0.5]{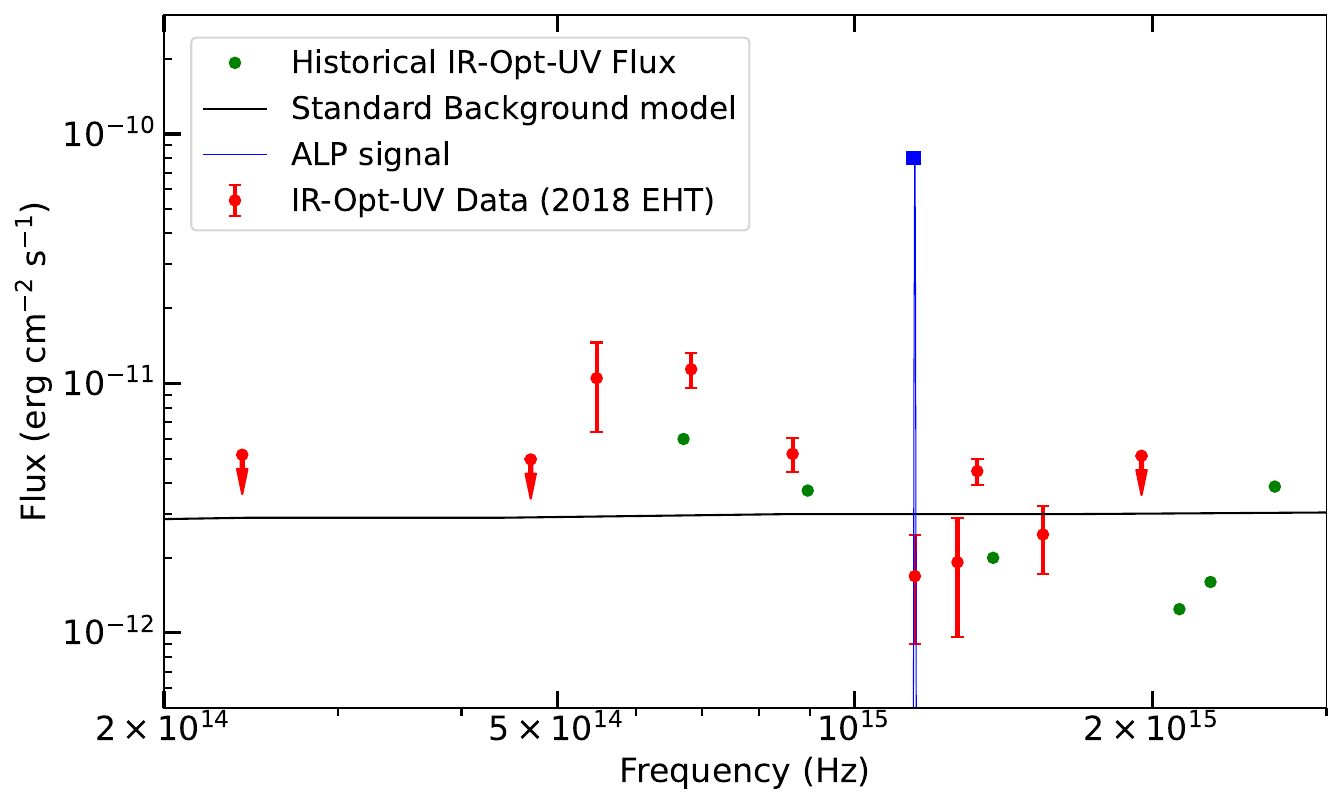}
\caption{The observed broadband SED of M87, measured during the April 2018 Event Horizon Telescope (EHT) campaign using various instruments, are shown in red points. 
The measurements, represented by red points with error bars, indicate observed flux with experimental uncertainties, while red downward arrows denote upper limits on flux estimates. 
The black line depicts the best-fit model (modified model A) applied to the 
entire dataset, as mentioned in \cite{EventHorizonTelescope-Multi-wavelengthscienceworkinggroup:2024xhy}. 
Additionally, the green dots represent the historical data in the IR-optical-UV 
frequency range from earlier observations of M87, compiled from studies such as \cite{Perola1980,Biretta,Despringe,Beuther:2006ad,Shi:2006vs,Perlman2003,Perlman:2007zc,Birettasparks,Marshall:2001de,Tan:2006vk}. 
{\it Bottom panel:} The IR-optical-UV frequency range which is highlighted by a box 
in the top panel, is shown in the bottom panel as a zoomed-in view. 
Here the ALP DM decay signal from M87 is shown by the blue line. Such a signal is 
estimated for $m_{a}=9.51$ eV with the value of $g_{a\gamma\gamma}$ set to the corresponding 
existing bound, i.e., $g_{a\gamma\gamma}=1.52\times 10^{-11} \, \rm GeV^{-1}$.}
\label{fig:The_Data_Plot}
\end{figure*}

%

Our analysis emphasizes the infrared, optical, and UV observations within 
M87's broad spectrum, from which we can constrain the ALP-photon coupling.

The dataset utilized in this study, is drawn from the campaign paper \cite{EventHorizonTelescope-Multi-wavelengthscienceworkinggroup:2024xhy} which details the second 
multi-wavelength observational campaign on M87 by the MWL WG. The measured spectral energy distribution (SED) data for M87 can be accessed via the data archive at \cite{EHTWG_web}. 
The MWL WG collaboration used state-of-the-art data reduction techniques to publish this dataset, covering an electromagnetic flux range from $10^{10}$ to $10^{27}$ Hz. These 2018 flux measurements are represented as red points in Fig.~\ref{fig:The_Data_Plot}. 

During this observational campaign, in the IR-Optical-UV region, 
four telescopes have been utilized for observations.
The Swift UV-Optical Telescope (UVOT) observed the M87 galaxy with an angular resolution of \(3''\), collecting data over a wavelength range of approximately \(192\)-\(547\) nm. 
The Kanata Telescope contributed data up to an upper flux limit at 634.5 nm and 1250 nm 
within a \(10'' \times 10''\) field of view. The UV Imaging Telescope (UVIT), part of the AstroSat mission, provided observations with two telescopes: one focused on the near-UV (2000–3000 Å) and visible (3200–5500 Å) bands, and the other on the far-UV (1300–1800 Å), offering upper flux limits with a large \(300''\) field of view. We will be using single data point obtained in the near-UV channel. The Hubble Space Telescope (HST), operating in the UV-optical range, measured the flux at a central wavelength of \(236\) nm but with a relatively small resolution of \(0.1''\). 
In total, there are 10 data points in the IR-Optical-UV bands, with seven having associated error bars for flux estimation.

The M87 SED are also informed by a series of earlier datasets. These older datasets include core emission measurements (ranging from millimeter to X-ray frequencies) \cite{Perola1980,Biretta,Despringe,Beuther:2006ad,Shi:2006vs,Perlman2003,Perlman:2007zc,Birettasparks,Marshall:2001de,Tan:2006vk}, MOJAVE VLBA observations at radio frequencies \cite{Lister}, Chandra X-ray data from 2009 \cite{2009ApJchandra}, Fermi-LAT data \cite{Fermi-m87}, HESS data \cite{HESS:2005wao}, VERITAS data \cite{Acciari:2008ah}, and MAGIC data \cite{MAGIC:2022piy}. The infrared-optical-UV frequency range of this older dataset, referred to in this paper as ``Historical data", is displayed as green points in Fig.~\ref{fig:The_Data_Plot}. 

The details of these observations are summarized in appendix \ref{sec:obs_data}.

\section{Methodology  \&  Results }\label{Methodology and  Results}

The contributions of the ALP DM signal at the observed frequencies 
(i.e., in IR, Optical and UV) are estimated 
 within the angular resolution of the telescopes, focused towards the center of M87. The angular resolution is determined by the specifications of each telescope for specific frequency windows across the electromagnetic spectrum. 
Comparing the ALP signal with the observed data, we derive bounds on the ALP-two-photon coupling $g_{a\gamma\gamma}$.  
In this work, we take two different approaches to derive these bounds on 
$g_{a\gamma\gamma}$. Below, we describe each approach 
and the corresponding constraints obtained 
in each case~\footnote{A discussion on the possible 
impact of systematic uncertainties on the results is provided in Appendix~\ref{Systematics}.}.


\subsection{Conservative approach}\label{Conservative_approach}

\begin{figure*}[ht!]
\centering
\includegraphics[width=13cm,height=9cm]{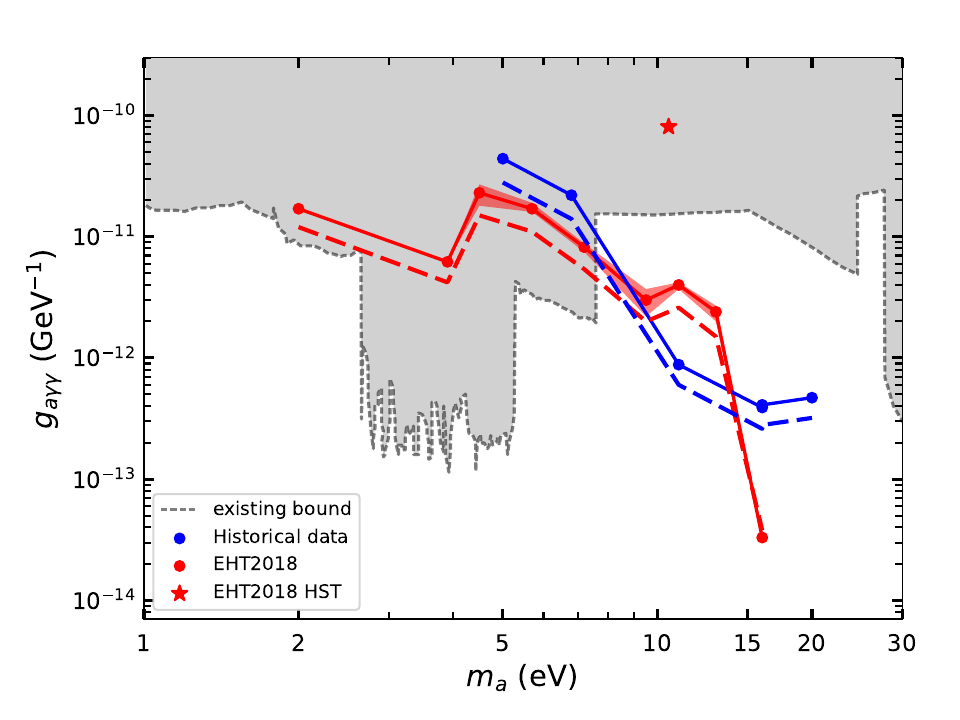}
\caption{Constraints on the axion photon coupling $g_{a\gamma\gamma}$ considering the conservative approach (described in Section~\ref{Conservative_approach}), 
is shown in the figure. 
The red and blue lines correspond to the 2018 observational campaign 
and the historical observational dataset of M87, respectively. 
The band around the constraint shown by the red solid line 
demonstrates the observational uncertainty in the data at relevant frequencies 
(see Table \ref{tab:observational_data}).
The single red star corresponds to the EHT(2018) HST observed data point. 
The dashed red and blue lines represent the constraints derived for an alternative 
NFW DM profile discussed in the text. 
The gray shaded region indicates the strongest existing bound 
in the $m_a - g_{a\gamma\gamma}$ plane \cite{Janish:2023kvi,Yin:2024lla,Regis:2020fhw,Grin:2006aw,
Carenza:2023qxh,Bernal:2022xyi,Wadekar:2021qae}.}
\label{fig:bounds_1}
\end{figure*}

This approach, serving as a conservative estimate, asserts that the DM signal 
must not exceed the observed data in any of the frequency bins that we are considering. 
This approach does not involve any modeling of the background, which makes it 
conservative in that regard. 
There are total ten data points (including observed data with error bars as well as 
upper-limits) from the EHT2018 in the IR-Optical-UV range. In addition, the historical dataset has six observed data points. For details, see Fig.~\ref{fig:The_Data_Plot} 
and Appendix \ref{sec:obs_data}. For a given ALP mass $m_a$ that corresponds to 
any of these observed frequencies, we obtained bound on $g_{a\gamma\gamma}$ 
requiring that the photon signal arising from the ALP decay 
(given in Eq.~\ref{ALP_signal}) does not exceed the observed data. 
The bounds obtained with this approach are shown in the 
$m_a - g_{a\gamma\gamma}$ plane in Fig.~\ref{fig:bounds_1}. 

As can be seen from Fig.~\ref{fig:bounds_1}, 
we are able to put constraints in the mass window (2-16) eV for the EHT(2018) dataset 
(shown by red solid and dashed curves) and in the mass window (5-20) eV for 
the historical dataset (blue solid and dashed curves). 
The constraints shown by the solid red and blue lines correspond to 
the NFW DM profile with the parameters obtained from \cite{HAWC:2023bti} 
as mentioned in Section \ref{Axion Signal}.  
In addition, we also show by the dashed red and blue lines 
the constraints that correspond to a NFW DM profile, 
but with different parameters: 
$\rho_{0} = 6.5\times 10^{7} \, M_{\odot}/{\rm kpc^{3}}$ , $r_{0} = 20$ kpc 
(obtained from \cite{Lacroix:2015lxa}). 
Apart from these, we also checked that considering a Burkert profile 
as reported in \cite{DeLaurentis:2022nrv}, with their best-fit parameters: 
$\rho_{0} = 4.7\times10^{-25} \, {\rm g/cm^3} = 6.94\times 10^{6} \, M_{\odot}/{\rm kpc^{3}}$ and $r_{0} = 91.2$ kpc, the constraints derived 
are weakened by a factor of $\sim1.8$ compared to that 
obtained using the former NFW profile. 
The rest of our discussion will be based on the former NFW profile parameters. In appendix~\ref{DM_profile_dependence} we present 
a more comparative analysis highlighting how the constraints vary with different choices of the dark matter profile for M87 galaxy.

We compare our bounds with the existing constraints 
(obtained from \cite{Janish:2023kvi,Yin:2024lla,Regis:2020fhw,Grin:2006aw,Carenza:2023qxh,Bernal:2022xyi,Wadekar:2021qae})~\footnote{The existing constraints are estimated from the data files corresponding to the relevant mass range, available on the website \cite{axionlimits_web}. The strongest limit on the coupling was identified and 
used as the upper boundary of the allowed parameter space.} 
that are shown as the gray-shaded region in the $m_a - g_{a\gamma\gamma}$ plane. 
We find that, our constraints can surpass the existing bounds 
on $g_{a\gamma\gamma}$ by an order of magnitude for the ALP mass range 
$8 \, {\rm eV} \lesssim m_a \lesssim 20 \, {\rm eV}$.  
Only for the single data from the HST observation, 
the constraint on the 
coupling (shown as red star) falls inside the existing bound 
due to the weak sensitivity of HST towards DM decay signal, owing to its 
much smaller angular resolution (see Table~\ref{tab:observational_data}). 

Our constraints on $g_{a\gamma\gamma}$ are comparatively stronger in the higher ALP mass range 
($8 \, {\rm eV} \lesssim m_a \lesssim 20 \, {\rm eV}$) than those at the lower ALP mass range 
($2 \, {\rm eV} \lesssim m_a \lesssim 8 \, {\rm eV}$). 
The photon flux collected by telescopes such as Kanata, Swift-UVOT, and IUE in the lower frequency range 
(corresponding to $2 \, {\rm eV} \lesssim m_a \lesssim 8 \, {\rm eV}$) are relatively higher than those 
in the higher frequency band (as evident from the bottom panel of Fig.~\ref{fig:The_Data_Plot}). 
Since the constraints on $g_{a\gamma}$ from ALP decay are derived considering these observed flux as the upper limits (under conservative assumptions), a larger observed flux in the low frequency range results in comparatively weaker bounds in the 
low ALP mass range. Also, independent of the observational data, the flux from ALP decay scales as $\propto g_{a\gamma}^2 m_a^3$. 
As a consequence, in the low-frequency regime (see again the bottom panel of Fig.~\ref{fig:The_Data_Plot}), 
a smaller axion mass leads to a suppressed signal (for a fixed coupling), and hence, 
a stronger coupling is required to yield a comparable flux, 
indicating that constraints on $g_{a\gamma}$ become progressively weaker for lower ALP masses.  
All these reasons inherently make the bounds on $g_{a\gamma}$ weaker at lower masses 
(i.e. $2 \, {\rm eV} \lesssim m_a \lesssim 8 \, {\rm eV}$ ) in comparison to the higher masses.


\subsection{$\chi^{2}$ fit constraints}

 In the second approach, the bounds on the coupling 
are derived considering a background fit model. We refer to this approach 
here as the $\chi^{2}$ fit constraints.
Based on the astrophysical background model that describes the observed data over the whole frequency range, we will incorporate photon emissions resulting from ALP decay to constrain the parameter space. 

\subsubsection{Constraints from standard background}\label{chisq_std_BG} 

\begin{figure*}[ht!]
\centering
\includegraphics[width=13cm,height=8cm]{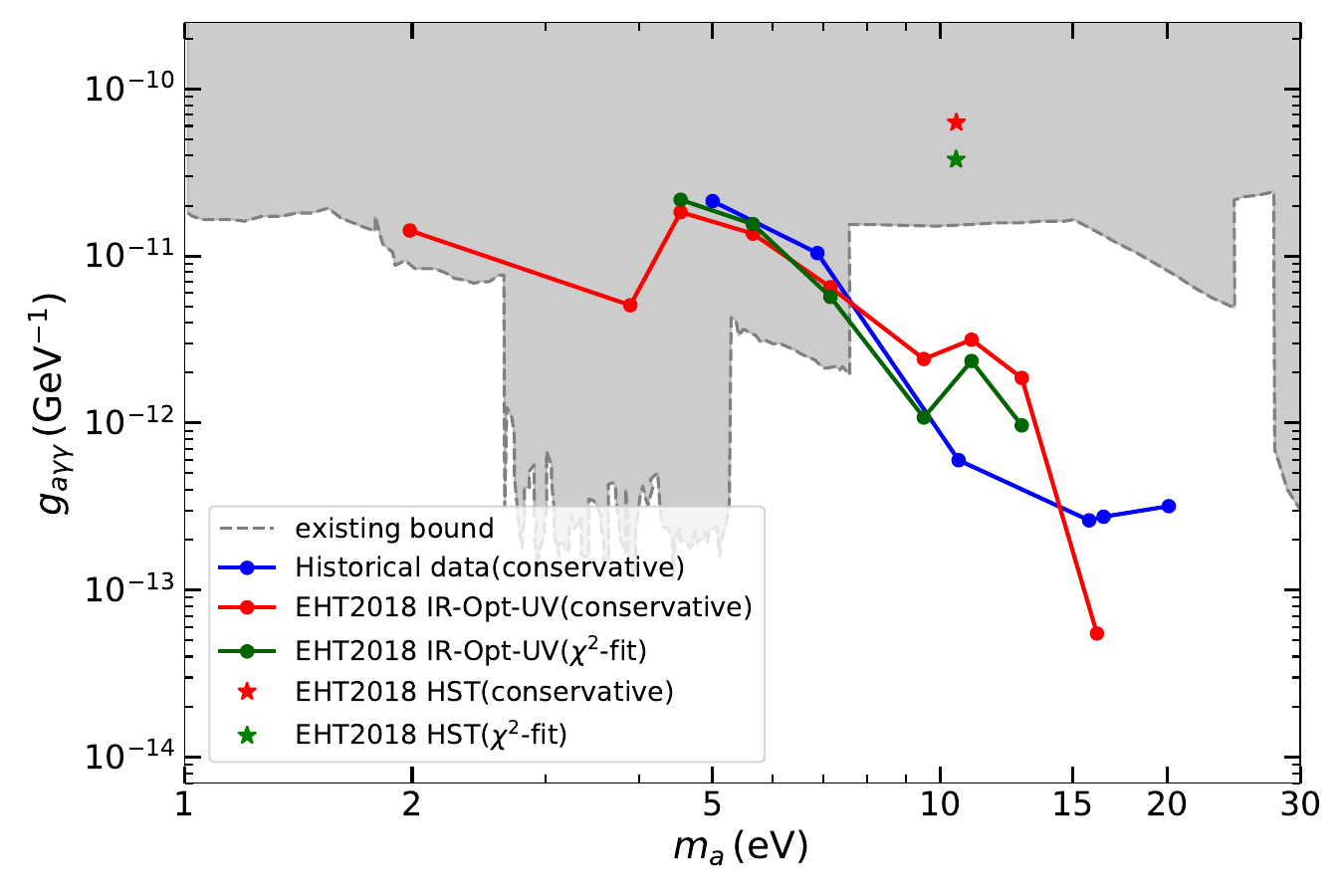}
\caption{Upper-limit on $g_{a\gamma\gamma}$ obtained (at 95$\%$ C.L.) 
considering a background model (modified model A from 
\cite{EventHorizonTelescope-Multi-wavelengthscienceworkinggroup:2024xhy}) 
is shown by the green line (see Section \ref{chisq_std_BG}).  
The result corresponds to the HST data is shown separately by green star. 
For comparison, constraints derived using the conservative approach 
are also presented: the red line corresponds to the EHT (2018) dataset, and the blue line corresponds to the historical dataset.}
\label{fig:bounds_2}
\end{figure*}

Given that the observed data are unevenly distributed across a wide frequency range, it is practical to fit a single model that accounts for the entire spectrum uniformly.
A comprehensive review of astrophysical models, including their parameters, is provided in MWL2018 \cite{EventHorizonTelescope-Multi-wavelengthscienceworkinggroup:2024xhy}. In this analysis, we adopt the modified Model A from \cite{EventHorizonTelescope-Multi-wavelengthscienceworkinggroup:2024xhy} as the standard astrophysical background.

 Now to estimate the coupling $g_{a\gamma\gamma}$, we perform a
 $\chi^2$ analysis
where $\chi^2$ is defined as:
\begin{equation}
\chi^2(m_a, g_{a\gamma\gamma}) = \sum_{i=1}^N \left(\frac{S_{\text{th}}^i - S_{\text{obs}}^i}{\sigma_{\text{obs}}^i}\right)^2 \, .
\label{eq:chisq}
\end{equation}
Here, \(S_{\text{th}}^i\) is the theoretical flux in the \(i\)-th frequency bin, computed as 
\(S_{\text{th}}^i = S_{\text{signal}}^i + S_{\text{background}}^i\). 
The $S_{\text{signal}}^i$ is the ALP DM signal computed using Eq.~\ref{ALP_signal}, and \(S_{\text{background}}^i\) 
is derived from the modified Model A.
\(S_{\text{obs}}^i\) is the observed flux, and \(\sigma_{\text{obs}}^i\) represents the associated observational uncertainty.

Here the summation runs over all frequency bins. For a fixed DM mass \(m_a\), 
the function \(\chi^2\) depends only on \(g_{a\gamma\gamma}\). 
The significance of a signal from ALP decay can then be computed using:
\begin{equation}
\Delta\chi^2 = \chi^2(g_{a\gamma\gamma}) - \chi^2_{\text{min}} \, .
\label{eq:del_chisq}
\end{equation}
For a 95\% confidence level (C.L.) and one degree of freedom, \(\Delta\chi^2 = 2.71\). 

The quantity $\chi^2_{\rm min}$, the minimum value of $\chi^2$, 
corresponds to the best-fitted background model 
\cite{EventHorizonTelescope-Multi-wavelengthscienceworkinggroup:2024xhy}
without any signal (that is, for a vanishing axion-photon coupling).
Taking into account the EHT(2018) whole dataset we first obtain $\chi_{min} ^{2}$,
then we add the DM signal {in the frequency bin corresponding to $m_a$} with varying $g_{a\gamma\gamma}$ 
and estimate the constraint on it within 95$\%$ C.L.

The constraint on $g_{a\gamma\gamma}$ obtained in this method is 
shown in Fig.~\ref{fig:bounds_2} by the green solid line, 
along with the conservative limit discussed in Section~\ref{Conservative_approach} (shown as red solid line) 
for comparison. The $\chi^2$-analysis of the entire dataset gives constraints in the mass range 
$4.3 \, {\rm eV} \lesssim m_a \lesssim 12.8 \, {\rm eV}$, of which the mass range 
$8 \, {\rm eV} \lesssim m_a \lesssim 12.8 \, {\rm eV}$ have almost an order of magnitude stronger constraints compared to 
the existing bounds.
The green star demonstrates the limit on the coupling for the frequency bin corresponding to HST data.

\begin{figure*}[ht!]
\centering
\includegraphics[width=9cm,height=6cm]{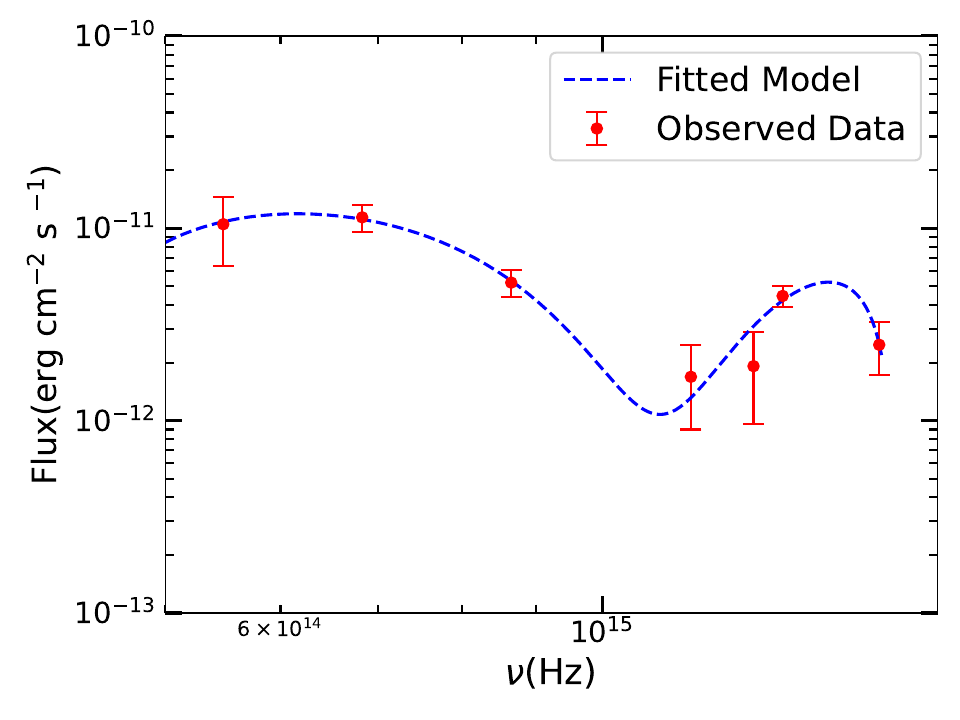} \\
\vspace{3mm}
\includegraphics[width=13cm,height=8cm]{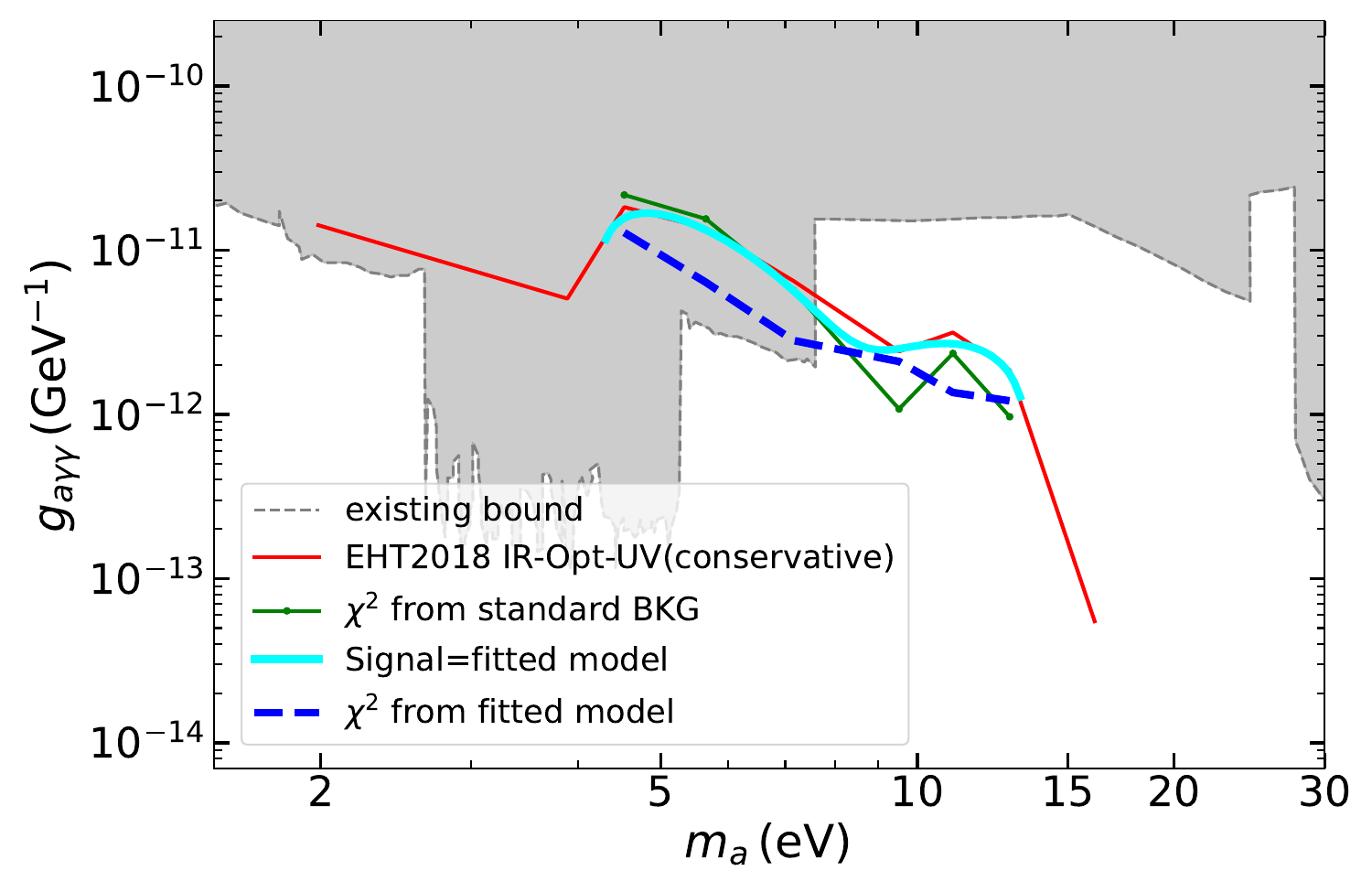}
\caption{{\it Top panel:} The blue dashed line represents the fit to the 2018 (EHT) data in the IR-Opt-UV frequency range (see Section \ref{chisq_fitted_bkg} for details). 
{\it Bottom panel:} The dashed blue line shows the constraint on $g_{a\gamma\gamma}$ 
derived (at 95$\%$ C.L.) using the fitted background model shown in the top panel. 
The solid cyan line corresponds to the constraint obtained 
assuming the fitted background model as the upper limit for the ALP DM decay signal. Constraints obtained using the standard background (BKG) model and the conservative approach 
are also presented for comparison.}
\label{fig:fit-constraint}
\end{figure*}

\subsubsection{Constraints from fitted background}\label{chisq_fitted_bkg} 

We also performed a fit to the infrared-optical-UV data points using a higher-degree polynomial model, as shown on the top panel of Fig.~\ref{fig:fit-constraint}. Due to the irregular distribution of these data points relative to the standard background, a conventional power-law model does not provide a good fit in this frequency range. Instead, we identified the best-fit curve 
for these 7 data points using a polynomial model. Based on this fitted model, the constraints on the $m_{a} - g_{a\gamma\gamma}$ 
parameter space are obtained.

The fitted model is expressed as:
\begin{equation}
    F(\nu) = a_0 + a_1 (\nu/\nu_0) + a_2 (\nu/\nu_0)^{2} + a_3 (\nu/\nu_0)^{3} + a_4 (\nu/\nu_0)^{4}
\end{equation}
where $\nu_{0}$ is a reference frequency chosen to be $10^{15}$ Hz. 
The fit yielded a $\chi^{2}/d.o.f = 0.98$ with the best-fit parameters: 
$a_0 = -2.05 \times 10^{-10} \, {\rm erg \, cm^{-2} \, s^{-1}}$, 
$a_1 = 9.92 \times 10^{-10} \, {\rm erg \, cm^{-2} \, s^{-1}}$, 
$a_2 = -1.605 \times 10^{-9} \, {\rm erg \, cm^{-2} \, s^{-1}}$, 
$a_3 = 1.08 \times 10^{-9} \, {\rm erg \, cm^{-2} \, s^{-1}}$ and 
$a_4 = -2.56 \times 10^{-10} \, {\rm erg \, cm^{-2} \, s^{-1}}$. 
This fitted model was then treated as the background for further analysis. 

Using this background model, the same $\chi^{2}$-fit method (discussed in Section~\ref{chisq_std_BG}
with a focus on the infrared-optical-UV part of the data)
was applied to constrain the coupling $g_{a\gamma\gamma}$ for a given $m_a$. 
In the bottom panel of Fig.~\ref{fig:fit-constraint}, 
the blue dashed line represents the 95$\%$ C.L. 
upper limit in the $g_{a\gamma\gamma}-m_{a}$ plane for the fitted background model. 
Additionally, the solid cyan line shows the constraints if the fitted model flux is directly interpreted as the 
upper bound of the ALP signal (i.e., without performing any $\chi^{2}$ analysis with the signal). 
For comparison, the conservative limit (discussed in Section~\ref{Conservative_approach}) as well as 
constraints obtained using the modified model A as background (discussed in Section~\ref{chisq_std_BG}) 
are also shown in the figure.

As shown in the figure, the fitted model provides slightly stronger constraints at the lower end of the mass range, whereas the standard background (the modified model A from \cite{EventHorizonTelescope-Multi-wavelengthscienceworkinggroup:2024xhy}) 
yields tighter constraints at the higher mass range.

\section{Conclusion}\label{conclusion} 

Due to its coupling with the SM photon, an axion-like particle (ALP) 
can decay into two photons. Assuming that ALPs account for the entire observed cold DM abundance of the Universe, their decay near the central region of galaxies, which is expected to be DM-rich, can give rise to observable line emission signals.   

Based on this, in this work we have constrained the ALP-two-photon coupling 
$g_{a\gamma\gamma}$ of the ALP DM with mass in the range 
$2 \, {\rm eV} \lesssim m_a \lesssim 20 \, {\rm eV}$, using the observation of 
M87 galaxy in infrared, optical and ultraviolet emission. 
For this we have used the dataset of the central region of M87, obtained by 
the EHT Multi-wavelength Science Working Group (MWL WG) during their 
2018 observational campaign, as well as the old dataset for the same region collected over the years. The recent 
data were collected by Swift-UVOT, HST, Astrosat-UVIT and Kanata telescope. 
{The historical data were collected by the spitzer space telescope and International Ultraviolet Explorer.}
Based on such data we exclude 
$g_{a\gamma\gamma} \gtrsim  10^{-11} \, \rm GeV^{-1} - 10^{-13} \, \rm GeV^{-1}$ 
in the above-mentioned mass window. 
{Fig.~\ref{fig:bounds_1} shows the results obtained in a conservative approach 
without assuming any background model.}
The $\chi^{2}$-fit analysis make the constraints stronger in the mass range 
$4 \, {\rm eV} \lesssim m_a \lesssim 13 \, {\rm eV}$ {which is presented in Fig.~\ref{fig:bounds_2} and Fig.~\ref{fig:fit-constraint}.} 
The better constraints are obtained by Swift-UVOT, Astrosat-UVIT and Kanata
telescope due to their wide field observations. 
Such constraints surpass the existing bounds on $g_{a\gamma\gamma}$ by 
an order of magnitude in the mass range 
$8 \, {\rm eV} \lesssim m_a \lesssim 20 \, {\rm eV}$. 
On the other hand, HST has a low sensitivity for the ALP DM decay signal 
due to its small angular resolution. 


The constraints can be improved through increasing the sensitivity of the observing telescopes and onboard spectrometers. One can improve the bounds by enhancing the data reduction maneuvers from the Swift-UVOT telescope in the upcoming future. 
Future telescopes like the LUVOIR and on-board spectrograph 
LUMOS \cite{luvoir_web} will be able to obtain higher sensitivity for the 
Infrared-optical-UV emission lines, and thus will be able to further improve 
the constraints on the ALP DM decay. UVEX~\cite{Kulkarni:2021tit} can serve as a future telescope in the UV frequency range.

\section*{Acknowledgements} 
AK acknowledges the hospitality of the Institut d’Astrophysique de Paris (IAP) 
where part of this work was done. PS receives support from the University Grants Commission, Government of India, through a Senior Research Fellowship.

\appendix

\section{Summary of observational data in the IR-Optical-UV range}
\label{sec:obs_data}
\begin{table}[ht!]
    \centering
    \begin{tabular}{@{}cccccc@{}}
    \toprule
        \(\nu\) (Hz) & Angular Scale [$\prime\prime$] & \( \nu S_{\nu} \)\(( \text{erg} \, \text{cm}^{-2} \, \text{s}^{-1} \)) &  Telescope/Observatory \\ 
        \midrule
        \(2.4\times 10^{14}\) & 10.0 & \(<5.18\times 10^{-12}\) & Kanata & \\
        \(4.7\times 10^{14}\) & 10.0 & \(<4.96\times 10^{-12}\) & Kanata & \\
        \(5.48\times 10^{14}\) & 3.0 & \(1.05\pm 0.41\times 10^{-11}\) & Swift-UVOT & \\
        \(6.83\times 10^{14}\) & 3.0 & \((1.14\pm 0.18)\times 10^{-11}\) & Swift-UVOT & \\
        \(8.65\times 10^{14}\) & 3.0 & \((5.22\pm 0.82)\times 10^{-12}\) & Swift-UVOT & \\
        \(1.15\times 10^{15}\) & 3.0 & \((1.69\pm 0.79)\times 10^{-12}\) & Swift-UVOT & \\
        \(1.27\times 10^{15}\) & 0.1 & \((1.92\pm 0.96)\times 10^{-12}\) & HST & \\
        \(1.33\times 10^{15}\) & 3.0 & \((4.45\pm 0.54)\times 10^{-12}\) & Swift-UVOT & \\
        \(1.55\times 10^{15}\) & 3.0 & \((2.48\pm 0.76)\times 10^{-12}\) & Swift-UVOT & \\ 
        \(1.95\times 10^{15}\) & 300.0 & \(<5.13\times 10^{-12}\) & AstroSat-UVIT & \\
        \bottomrule
    \end{tabular}
    \caption{This table summarizes the observed data for infrared, optical and ultraviolet emissions from the central region of the M87 galaxy. These data were gathered during 2018 campaign (by MWL WG) by various telescopes such as Swift-UVOT, Kanata, Astrosat-UVIT, HST at the frequencies listed in the first column. The angular resolutions used by different telescopes are provided in arc-seconds($\prime\prime$) corresponding to their respective working frequencies. The data collected by the Swift-UVOT and HST include the experimental errors in flux measurements, while the data collected by Kanata and Astrosat-UVIT only provide upper limits on the flux. Data in this table are reproduced from \cite{EventHorizonTelescope-Multi-wavelengthscienceworkinggroup:2024xhy} under CC BY 4.0.}
    \label{tab:observational_data}
\end{table}
The data central to this study, focusing on infrared, optical, and ultraviolet emissions from the central region of the M87 galaxy, are summarized in the table~\ref{tab:observational_data}. The multi wavelength dataset of M87 can be obtained from the data table of the MWL WG published article \cite{EventHorizonTelescope-Multi-wavelengthscienceworkinggroup:2024xhy}. The table~\ref{tab:observational_data} represents the working frequencies of the telescopes along with their resolutions and measured fluxes. On the other hand, we also incorporated a historical dataset, collected over several years \cite{Perola1980,Biretta,Despringe,Beuther:2006ad,Shi:2006vs,Perlman2003,Perlman:2007zc,Birettasparks,Marshall:2001de,Tan:2006vk}, which is represented by green dots in the Fig.~\ref{fig:The_Data_Plot} (not shown in a table). {The historical data in the infrared-optical range were collected by spitzer space telescope with telescopic resolution in the ballpark of $\simeq 1 ''$ and in the ultraviolet region the data were  obtained by the International Ultraviolet Explorer (IUE) with a resolution of $\simeq 10''$.}
\begin{figure*}
\centering
\includegraphics[width=13cm,height=9cm]{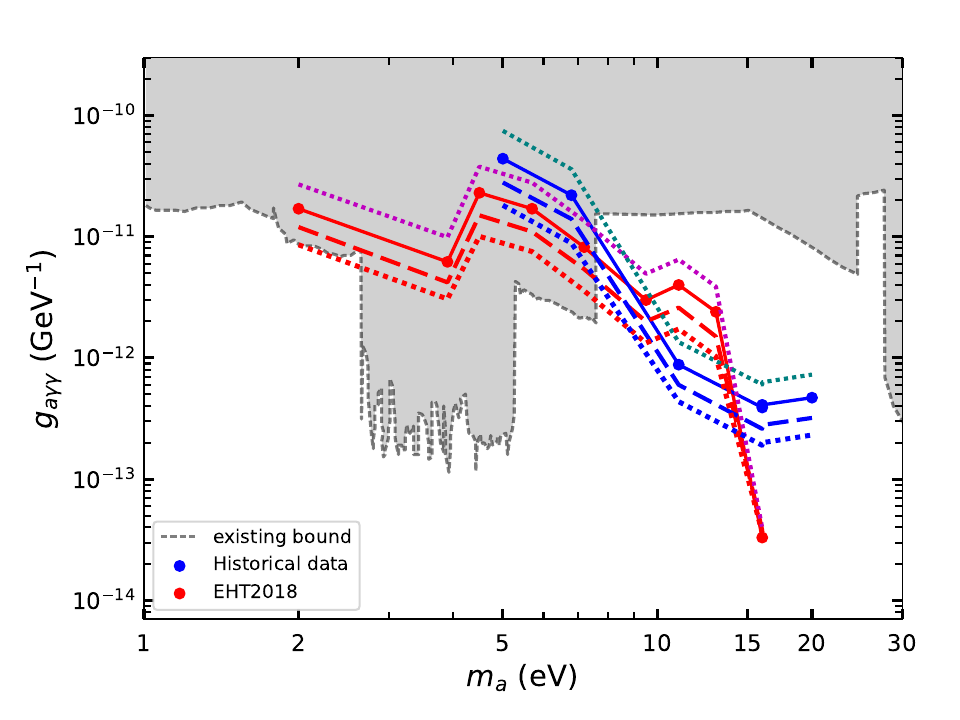}
\caption{The plot illustrates the variation of the constraints 
with different assumptions for the DM halo profile of M87, 
using both the new and historical dataset. 
Solid lines represent an NFW profile as adopted in \cite{HAWC:2023bti}, 
while dashed lines correspond to an alternative NFW profile from \cite{Lacroix:2015lxa}. They are same as the ones shown in Fig.~\ref{fig:bounds_1}. Constraints derived using a contracted NFW (cNFW) profile with 
$\gamma = 1.25$ are depicted with dotted red and blue lines. 
Bounds corresponding to a cored Burkert profile are shown 
with dotted purple and green lines.}
\label{fig:bounds_diff_profiles}
\end{figure*}

\section{Dependence on DM halo profile} \label{DM_profile_dependence}

Here we elaborate the dependence of our results on the choices of 
DM halo profile considered for the M87 galaxy. 
We explore several DM profiles. 
As discussed in the main text, we consider the NFW profile from the recent 
study~\cite{HAWC:2023bti} (Eq.~\ref{eq:NFW_profile}), 
and an alternative NFW profile from \cite{Lacroix:2015lxa}, 
both assuming a typical inner slope parameter of $\gamma = 1$. 
The solid and dashed (red and blue) lines, respectively, 
depict the limits derived using these two 
NFW profiles and are same as the ones shown in Fig.~\ref{fig:bounds_1}. 
Note that the profile from \cite{HAWC:2023bti} is used to obtain 
our main results.

In addition, we also adopt the NFW profile from 
\cite{Lacroix:2015lxa} with a steeper inner slope 
$\gamma = 1.25$~\cite{Fermi-LAT:2017opo}, leading 
to a contracted NFW (cNFW) profile. For this we choose 
$r_0 = 20$ kpc (same as in the case of $\gamma = 1$) 
and set the density normalisation $\rho_0$ in such a way that 
the total mass of the galaxy within 50 kpc remains same 
as in the case of $\gamma = 1$. 
The corresponding constraints obtained using 
this cNFW profile are represented in 
Fig.~\ref{fig:bounds_diff_profiles} by dotted red and blue lines.

Incorporating a central DM spike on top of the NFW profile 
from \cite{Lacroix:2015lxa} (with $\gamma = 1$) 
leads to only a marginal change in the results 
($\lesssim1\%$) compared to that obtained for the NFW profile. 
This is mainly due to the fact that the DM decay signal is mostly 
sensitive to the overall spread of the DM distribution. 
As this impact is negligible, the corresponding result has not been 
displayed in the figure.

Additionally, we include the constraints based on a cored 
Burkert profile \cite{DeLaurentis:2022nrv}, 
shown as dotted purple and green lines in Fig.~\ref{fig:bounds_diff_profiles}. 
The specific values of the profile parameters 
such as $\rho_0$ and $r_0$ correspond to their best-fit values 
from \cite{DeLaurentis:2022nrv} (also provided in the main text).

\section{Systematic Uncertainty}\label{Systematics}

The dataset published by the MWL WG, which is utilized in this study, does not explicitly detail the systematic uncertainties arising from detector calibration. However, several systematic factors could influence the results.
In this section, we will discuss a few of them.

\subsection{Background modeling}
In our $\chi^2$ analysis in Sec.~\ref{chisq_std_BG}, we employed the modified Model A as the background model, as presented in \cite{EventHorizonTelescope-Multi-wavelengthscienceworkinggroup:2024xhy}. If we instead consider alternative background models, such as Model A or Model B from that reference, which provide the best fit to the overall spectrum, the resulting constraints on the coupling will exhibit only minimal variations. The table~\ref{tab:diff_bkg_model_constraints} 
below illustrates the changes in the results 
(i.e. upper limit on $g_{a\gamma\gamma}$) when using Model A and Model B in comparison to the modified Model A.
\begin{table}[h]
    \centering
    \renewcommand{\arraystretch}{1.3} 
    \setlength{\tabcolsep}{10pt}      
    \begin{tabular}{cccc}
        \toprule
        Mass Bin & Modified Model A & Model A & Model B \\
        $m_a$(eV) & \textbf{$g_{a\gamma\gamma}$ ($\times 10^{-12}\,GeV^{-1}$)} & \textbf{ $g_{a\gamma\gamma}$($\times 10^{-12}\,GeV^{-1}$)} & \textbf{ $g_{a\gamma\gamma}$($\times 10^{-12}\,GeV^{-1}$)} \\
        \midrule
        4.53  & \(21.7 \) & \(22.8 \) & \(23.2\) \\
        5.65   & \(15.5\) & \(16.4\) & \(16.6 \) \\
        7.16  & \(5.69\) & \(6.95\) & \(7.14\) \\
        9.51   & \(1.07\) & \(1.73\) & \(1.80\) \\
        11.00 & \(2.35\) & \(3.20\) & \(3.20\) \\
        12.82 & \(0.97\) & \(1.58\) & \(1.57\) \\
        \bottomrule
    \end{tabular}
    \caption{The table presents the upper bounds on the photon-ALP coupling $g_{a\gamma\gamma}$ across various mass bins, derived using three distinct background models, taken from \cite{EventHorizonTelescope-Multi-wavelengthscienceworkinggroup:2024xhy}.}
    \label{tab:diff_bkg_model_constraints}
\end{table} 

As discussed in Sec. \ref{chisq_fitted_bkg}, we have also modeled the background by fitting the data with a fourth-order polynomial, as it provides the best fit with the lowest $\chi^2$/d.o.f. compared to polynomials of other orders. 
For example, considering a third-order polynomial, 
$F(\nu) = b_0 + b_1 (\nu/\nu_0) + b_2 (\nu/\nu_0)^{2} + b_3 (\nu/\nu_0)^{3}$, where $\nu_{0}$ is a reference frequency chosen to be $10^{15}$ Hz 
with fitting parameters: 
$b_0 = 8.99 \times 10^{-11} \, {\rm erg \, cm^{-2} \, s^{-1}}$, 
$b_1 = -2.14 \times 10^{-10} \, {\rm erg \, cm^{-2} \, s^{-1}}$, 
$b_2 = 1.73 \times 10^{-10} \, {\rm erg \, cm^{-2} \, s^{-1}}$ and 
$b_3 = -4.64 \times 10^{-11} \, {\rm erg \, cm^{-2} \, s^{-1}}$, results in $\chi^2$/d.o.f. of $4.01$. 
Using this third-order polynomial would alter the constraints by no more than $5\%$, but it would increase $\chi^2$/d.o.f., leading to a poorer fit to the data. 

\subsection{Detector calibration}
Systematic errors in detector calibration arise from factors such as inaccuracies in converting count rates to flux, changing sensitivity over time (aging effects), and the Point Spread Function (PSF). In observations of M87, the PSF influences how light from its extended structure is spread and blurred by the telescope, affecting measurements of surface brightness, flux accuracy, and the ability to resolve fine details in both the central and outer regions.

The choice of background regions can also contribute to systematic effects, especially in cases where diffuse emissions from the galaxy or variations in the sky background are present. Furthermore, in observations of bright sources, coincidence loss occurs when multiple photons arrive in the same detector region within a short time, causing non-linear count rates.

For Swift-UVOT calibration, a systematic error of approximately 2.3$\%$ is assumed for individual UVOT photometric measurements, as reported in calibration studies by \cite{Poole:2007xi}. The analysis by \cite{Breeveld:2011eb} indicate a $\sim 1 \%$ per year decline in UVOT sensitivity across all filters. More recently, \cite{Zhou:2025jlq} claimed that for the wavelength range 
$2800\,\text{\AA} < \lambda < 4000\,\text{\AA}$
 , the systematic uncertainty in the extraction of the source data 
is about $11.2\%$.

AstroSat-UVIT, a space-based telescope operating in the near-UV region, maintains systematic uncertainties below $\lesssim 15\%$ \cite{Tandon_2017,2020AJ....159..158T}. Meanwhile, the ground-based Kanata telescope, subject to atmospheric and instrumental systematics, typically experiences uncertainties in the range of $\sim(3-5)\%$ \cite{Ikejiri:2011ek,Sasada:2011yh,Itoh:2013rj}.

Old infrared observations (for the historical data) of M87 conducted with the Spitzer telescope had a photometric uncertainty of approximately $\sim 3\%$ \cite{Bohlin:2014nwa,SpitzerIRAC}. Similarly, ultraviolet observations by the IUE exhibited an uncertainty in the range of $\sim(3-5)\%$ \cite{Massa:1998nr,2018AJ....155..162B}. For a more detailed discussion and rigorous analysis of these uncertainties, 
we refer the reader to the cited references.

As long as these uncertainties stay below the level of the statistical uncertainties, their influence on the results appears to be limited, with only minor variations observed in the constraints across different mass bins.

\subsection{DM profile uncertainty:}
The variations of the results depending on the choice of dark matter distribution profiles for the M87 galaxy is detailed in appendix~\ref{DM_profile_dependence}.

\bibliographystyle{JHEP}
\bibliography{References}
\appendix

\end{document}